\documentclass[sigconf]{acmart}

\copyrightyear{2026}
\acmYear{2026}
\setcopyright{cc}
\setcctype{by}
\acmConference[UbiComp Companion '26]{Companion of the 2026 ACM International Joint Conference on Pervasive and Ubiquitous Computing}{October 11--15, 2026}{Shanghai, China}
\acmBooktitle{Companion of the 2026 ACM International Joint Conference on Pervasive and Ubiquitous Computing (UbiComp Companion '26), October 11--15, 2026, Shanghai, China}
\acmDOI{10.1145/3798063.3837839}
\acmISBN{979-8-4007-2533-3/2026/10}

\title[SmartFlex Adaptive Lumbar Support]{SmartFlex: An Adaptive Lumbar Support System Based on Posture Recognition and Air Bag Array}

\author{Ben Xiaolu Huang}
\email{benhsfls@gmail.com}
\affiliation{%
  \institution{Shanghai Foreign Language School Affiliated to SISU}
  \city{Shanghai}
  \country{China}
}

\begin{document}

\begin{abstract}
Low back pain (LBP) is a leading cause of disability worldwide and affects populations ranging from working adults to students with prolonged sitting habits. Conventional lumbar support belts are generally static and non-adaptive, which limits their ability to accommodate dynamic postural changes and individualized comfort requirements.

This paper presents SmartFlex, an intelligent wearable lumbar support system that integrates real-time posture recognition with an adaptive air bag array. The system uses a JY901S gyroscope sensor to detect user posture and a lightweight TinyML neural network deployed on an Arduino R4 UNO to process posture data at the edge. Based on the recognized posture state, a closed-loop pneumatic control system dynamically inflates or deflates 14 distributed air bags through four independent micro air pumps to provide targeted biomechanical support. Evaluation results show that SmartFlex achieves over 94\% posture recognition accuracy and generates corresponding pressure-control commands with a sensing-to-command delay of less than 120~ms. The pneumatic system operates within a calibrated pressure range of 15--85~kPa. A user study with 20 participants produced a 4.5/5 rating for support effectiveness, suggesting that adaptive wearable support may improve daily sitting comfort and reduce lumbar fatigue.
\end{abstract}



\maketitle

\section{Introduction}
Low back pain remains a critical public health challenge. In 2020, it affected approximately 619 million people globally, and its prevalence is projected to reach 843 million by 2050~\cite{who2023lowbackpain}. As a major contributor to years lived with disability, LBP also produces substantial economic costs through reduced productivity and increased medical expenditure. Although LBP has traditionally been associated with aging and physically demanding labor, prolonged sitting, intensive study routines, and frequent electronic-device use increasingly expose adolescents and young adults to posture-related spinal stress.

Existing non-invasive interventions, including extensible and inextensible lumbosacral orthoses, face a persistent trade-off between stability and comfort. Rigid systems can provide stronger trunk stabilization but often restrict range of motion and reduce long-term wearability; flexible systems are more comfortable but may provide insufficient targeted support under changing postural demands~\cite{cholewicki2010comparison,han2022insilico,morrisette2014randomized}. This limitation motivates a support system that can adapt its mechanical response to the user's posture rather than applying a constant static force.

To address this gap, SmartFlex combines wearable sensing, embedded machine learning, and pneumatic actuation. The system continuously monitors posture, estimates the support required by different lumbar and abdominal regions, and redistributes pressure across an air bag array. Its design objective is to provide dynamic, personalized support while preserving comfort during everyday sitting and movement.

\section{Technical Approach}
\subsection{System Architecture}
\begin{figure}[t]
  \centering
  \includegraphics[width=\linewidth]{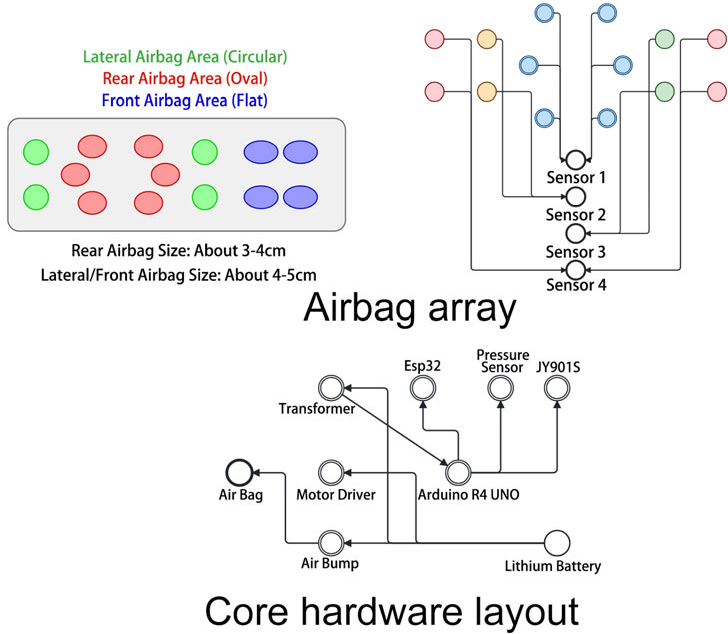}
  \caption{System architecture detailing the 14-airbag array distribution and core hardware layout.}
  \Description{The figure combines an air bag distribution diagram with a simplified hardware layout. It shows lateral, rear, and front air bag zones, four pressure sensors, an Arduino R4 UNO, an ESP32 module, a JY901S sensor, motor drivers, micro air pumps, and a lithium battery.}
  \label{fig:architecture}
\end{figure}

The SmartFlex hardware follows a layered modular architecture, as shown in Figure~\ref{fig:architecture}. The sensing layer uses a JY901S gyroscope positioned near the lumbar spine to capture pitch and roll information at a sampling frequency of 100~Hz. These spatial measurements are transmitted to an Arduino R4 UNO, which functions as the central embedded controller.

The actuation layer consists of four independent micro air pump systems connected to a biomechanically mapped array of 14 thin-film air bags. Six posterior air bags are placed near the L3--L5 region to counteract extensor muscle fatigue. Four lateral air bags support biomechanical balance during lateral flexion, while four anterior air bags assist the maintenance of physiological lordosis during sitting and forward flexion.

\subsection{Intelligent Control and TinyML Integration}
To achieve low-latency adaptive support without relying on continuous cloud computing, a lightweight Multilayer Perceptron neural network was trained on human posture data and optimized using TensorFlow Lite for edge deployment on the Arduino R4 UNO. The network topology consists of an input layer receiving standardized pitch and roll values, two hidden layers with 24 and 16 neurons using ReLU activation and batch normalization, and an output layer with four neurons corresponding to four general posture classes: (1) neutral posture; (2) forward flexion; (3) backward extension; and (4) lateral flexion.

When lateral flexion is recognized, the sign of the roll angle is used to distinguish leftward and rightward bending. This allows the four-class model to generate five practical control states: neutral, forward flexion, backward extension, left lateral flexion, and right lateral flexion.

To adapt to the embedded system's memory limitations, post-training quantization was applied to convert model parameters from floating-point values to 8-bit integers, reducing runtime computational overhead and enabling fixed-point arithmetic for inference.

Once a posture transition is classified, a separate posture-to-pressure mapping module establishes the target execution profile for the four pneumatic zones. The target pressure of zone $i$ is defined as
\begin{equation}
  P_i^{*} = 15 + (85-15)m_i,
  \label{eq:target-pressure}
\end{equation}
where $0 \leq m_i \leq 1$ is the support coefficient assigned according to the recognized posture and bending direction. This mapping constrains all target pressures to the calibrated operating range of 15--85~kPa. The 15~kPa lower value represents the baseline pneumatic state, while higher values provide additional support to the principal zone associated with the detected posture.

During forward flexion, the anterior zone $p_2$ receives increased pressure. During backward extension, the posterior zone $p_1$ is emphasized. During lateral flexion, the roll direction determines whether the left zone $p_3$ or the right zone $p_4$ receives additional pressure. A transition factor, $\alpha=0.7$, is applied to update the pressure command gradually:
\begin{equation}
  P_i^{\mathrm{cmd}}(t+1) = P_i^{\mathrm{cmd}}(t) + \alpha\left(P_i^{*} - P_i^{\mathrm{cmd}}(t)\right).
  \label{eq:pressure-update}
\end{equation}
This incremental update reduces abrupt changes in the commanded pressure during posture transitions.

\begin{figure}
  \centering
  \includegraphics[width=\linewidth]{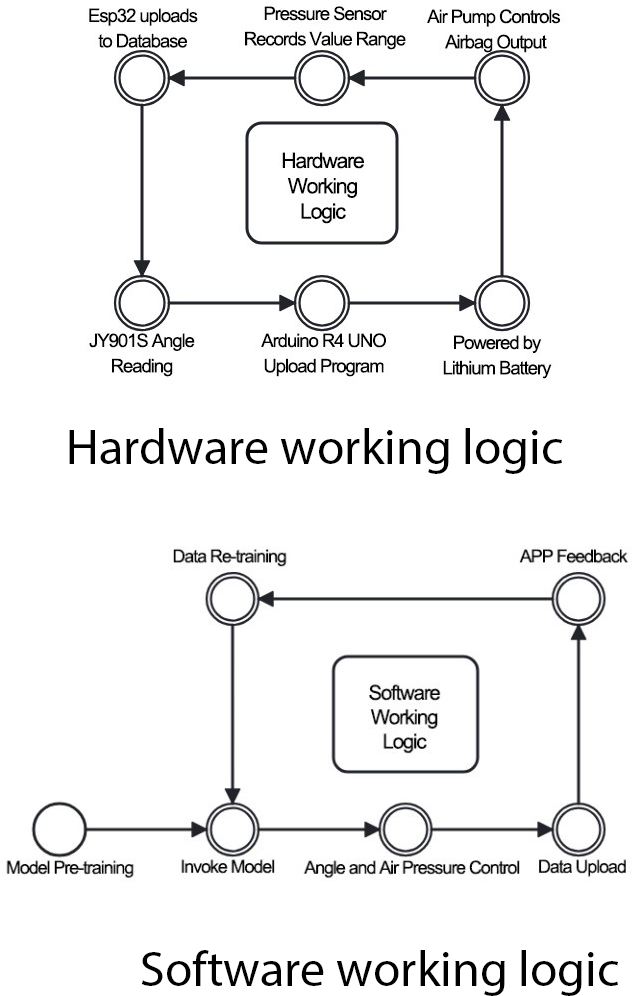}
  \caption{Closed-loop execution cycle and software working logic.}
  \Description{The figure shows two flow diagrams. The hardware loop links the JY901S angle reading, Arduino R4 UNO program, lithium battery, air pump output, pressure sensor feedback, and ESP32 database upload. The software loop links model pre-training, model invocation, angle and pressure control, data upload, application feedback, and data retraining.}
  \label{fig:control}
\end{figure}

A closed-loop feedback mechanism utilizes four pressure sensors to monitor the pneumatic states of the four zones. The controller compares the measured pressure with the commanded value and activates the corresponding inflation or pressure-release path when the deviation exceeds a preset dead-zone threshold. When the measured pressure is sufficiently close to the command, the current pneumatic state is maintained to reduce unnecessary switching and mechanical wear. Concurrently, an ESP32 module transmits telemetry data through the MQTT protocol to a cloud service connected to a MongoDB database for long-term storage and analysis. Wireless data transmission operates independently of the local posture-recognition and pressure-control loop. Figure~\ref{fig:control} summarizes the hardware and software execution logic.

\section{Prototype and Evaluation}
\subsection{Prototype Implementation}
The SmartFlex prototype was fabricated by attaching the 14-air bag array to the inner surface of a breathable elastic support belt using high-strength thermal adhesive. The micro air pumps, motor drivers, and microcontrollers were consolidated on a custom 3D-printed mounting platform designed in Fusion 360 and secured to the exterior of the belt. This arrangement supports heat dissipation while preserving the belt's ergonomic contour. The untethered prototype is powered by a 12~V lithium battery with an integrated voltage converter.

\begin{figure}
  \centering
  \includegraphics[width=\linewidth]{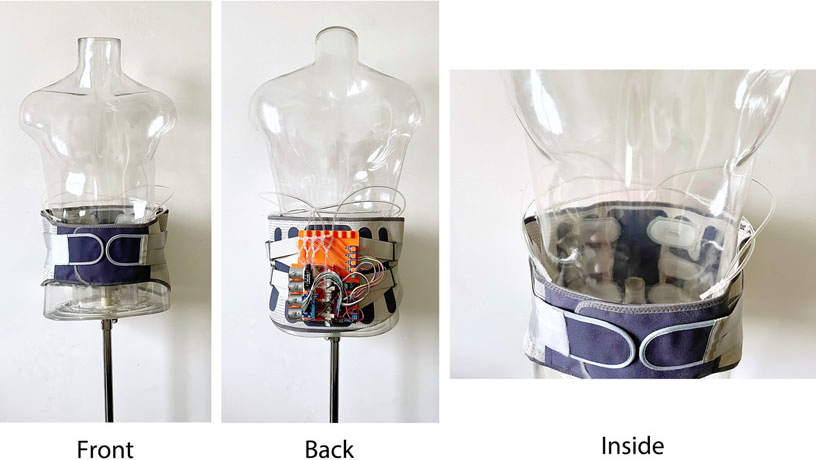}
  \caption{SmartFlex prototype showing the front, back, and inside configurations.}
  \Description{The figure contains three photographs of the prototype mounted on a transparent mannequin. The front view shows the belt and air bag placement, the back view shows the external electronics platform, and the inside view shows the contact surface and tubing.}
  \label{fig:prototype}
\end{figure}

\subsection{Performance and User Evaluation}
We conducted laboratory testing and human-centered evaluations to assess system viability. In laboratory trials involving 10 participants executing repeated static and dynamic movements, the TinyML model achieved over 94\% classification accuracy across the four predefined posture classes.

The closed-loop pneumatic system maintained stable operation across the four zones while regulating pressure within the calibrated 15--85~kPa range. A $\pm 2.5$~kPa dead-zone threshold was used by the controller to determine whether inflation, pressure release, or pressure holding was required.

The measured sensing-to-command delay was less than 120~ms during the tested posture transitions. This interval covered posture-data acquisition, TinyML inference, posture-to-pressure mapping, and generation of the corresponding pneumatic-control command.

\begin{figure}[t]
  \centering
  \includegraphics[width=\linewidth]{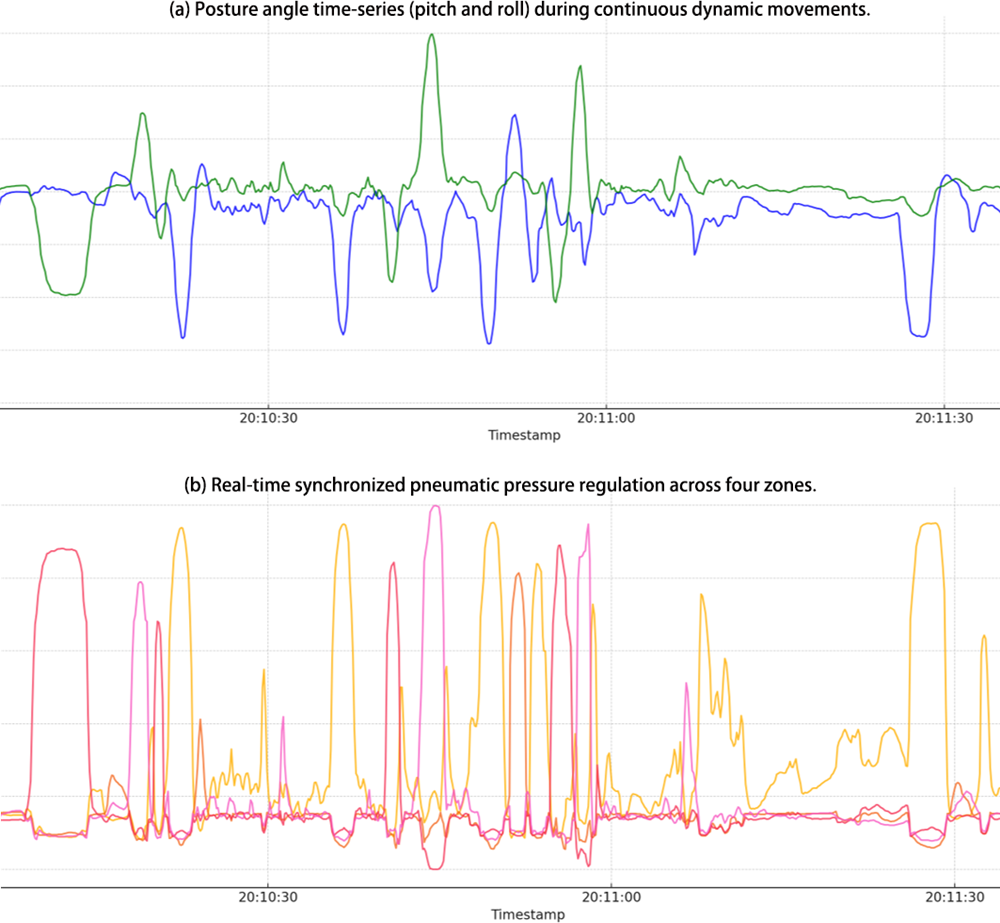}
  \caption{System dynamic tracking performance during repeated motion trials.}
  \Description{The figure contains two time-series plots. The upper plot shows pitch and roll angle variation over time during continuous movement. The lower plot shows synchronized pressure changes across four pneumatic zones over the same time interval.}
  \label{fig:dynamic}
\end{figure}

Quantitative time-series analysis demonstrated a clear temporal relationship between user posture angles and pressure adjustments. When a user transitioned into forward flexion, the pitch angle increased significantly, and the system generated an increased pressure command for the anterior abdominal zone, $p_2$. During backward extension, the posterior lumbar zone, $p_1$, provided the primary response. During lateral bending, the roll angle changed according to the bending direction. The corresponding lateral zone, $p_3$ or $p_4$, then executed targeted pressure modulation. The synchronized curves in Figure~\ref{fig:dynamic} show that the four pneumatic zones responded selectively to the measured posture changes.

To assess practical utility, a user-experience study was conducted with 20 volunteers, including individuals experiencing mild lumbar discomfort. Participants evaluated the device across simulated daily activities, including desk work, standing, bending, and ordinary movement.

\clearpage
\begin{figure}[t]
  \centering
  \includegraphics[width=\linewidth]{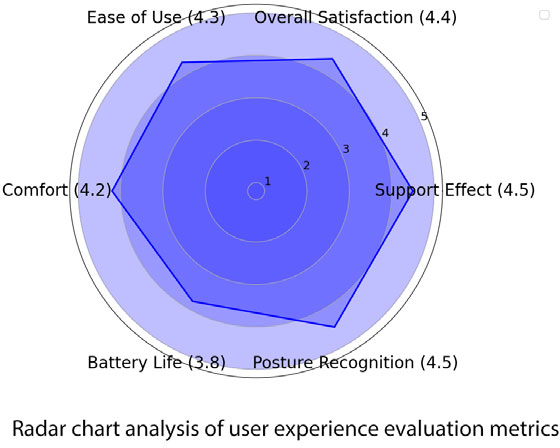}
  \caption{Radar chart detailing user-experience evaluation across six key performance metrics.}
  \Description{The radar chart reports user ratings for support effect, posture recognition, battery life, comfort, ease of use, and overall satisfaction on a five-point scale.}
  \label{fig:radar}
\end{figure}

The evaluation results were positive. The six mean ratings shown in Figure~\ref{fig:radar} were:
\begin{itemize}
  \item Support Effect: 4.5/5;
  \item Posture Recognition: 4.5/5;
  \item Overall Satisfaction: 4.4/5;
  \item Ease of Use: 4.3/5;
  \item Wearable Comfort: 4.2/5;
  \item Battery Life: 3.8/5.
\end{itemize}
Notably, 85\% of participants reported evident fatigue relief during prolonged desk work. Battery Life received the lowest of the six ratings, highlighting a clear direction for future hardware and power-management optimization.

\subsection{Ethics, Consent, and Safety Monitoring}
The evaluation protocol received school-level ethical review and was conducted under the supervision of the project advisor. Before participation, all volunteers were informed of the study purpose, wearing procedure, data collection process, and possible temporary discomfort, and provided informed consent. Participation was voluntary, and participants could pause or withdraw from testing at any time.

Before each session, the belt, air bags, tubing, electrical connections, and pneumatic components were inspected. Testing was supervised throughout, pressure commands were constrained to the calibrated 15--85~kPa range, and the device could be immediately depressurized and powered off if a participant reported discomfort or if abnormal operation was observed.

\subsection{Limitations}
Despite the promising functional evaluation, several limitations exist in the current prototype. First, the TinyML posture-classification algorithm may occasionally produce brief misidentifications when processing complex and highly rapid compound-motion sequences, primarily because these movements contain overlapping pitch and roll characteristics.

Second, the mechanical volume and power consumption of the four independent micro air pump systems constrain the device's overall portability and battery endurance.

Finally, further development of the data platform and mobile interface could enable more refined pressure-control strategies, interactive posture feedback, and long-term analysis of users' postural patterns.

\section{Conclusion}
The SmartFlex Belt demonstrates how ubiquitous sensing and embedded machine learning can transform traditional passive wearables into active, intelligent support systems. By integrating real-time posture recognition with four-zone pneumatic control, the system dynamically redistributes pressure across the posterior, anterior, and lateral regions of the waist.

The prototype achieved over 94\% posture-classification accuracy, generated pneumatic-control commands within 120~ms, and operated across a calibrated pressure range of 15--85~kPa. User evaluation results also indicated positive perceived support, posture responsiveness, wearable comfort, ease of use, and overall satisfaction.

Future iterations will focus on hardware miniaturization, pump and battery optimization, improved recognition of compound movements, refined pressure-control strategies, and the integration of a mobile application to provide interactive posture guidance and long-term data tracking.

\bibliographystyle{ACM-Reference-Format}
\bibliography{references}

\end{document}